\documentclass[letterpaper,showpacs,preprintnumbers,amsmath,amssymb,nofootinbib,twocolumn,superscriptaddress,notitlepage,prl]{revtex4-1}

\usepackage{graphicx}
\usepackage[usenames,dvipsnames]{color}
\usepackage[
colorlinks=true,       % false: boxed links; true: colored links
linkcolor=red,          % color of internal links
citecolor=blue,        % color of links to bibliography
filecolor=magenta,      % color of file links
urlcolor=cyan  
]{hyperref}

\usepackage{amsmath}
\usepackage{bbm}
\usepackage{amsfonts}
\usepackage{amssymb}
\usepackage{latexsym}
\usepackage{graphicx}
\usepackage[english]{babel}
\usepackage{multirow}
\usepackage{float}
\usepackage{url}
\usepackage{slashed}
\usepackage{xcolor}

\newcommand{\be}{\begin{equation}}
\newcommand{\ee}{\end{equation}}
\newcommand{\ba}{\begin{array}}
\newcommand{\ea}{\end{array}}
\newcommand{\bea}{\begin{eqnarray}}
\newcommand{\eea}{\end{eqnarray}}

\usepackage{ulem,fancyvrb}
\usepackage{xcolor}

\begin{document}

\title{On-shell mediator dark matter models and the Xenon1T anomaly}

\author{Mingxuan Du} 
\affiliation{Department of Physics, Nanjing University, Nanjing 210093, China}

\author{Jinhan Liang} 
\affiliation{Department of Physics, Nanjing University, Nanjing 210093, China}

\author{Zuowei Liu} %\email{zuoweiliu@nju.edu.cn}
\affiliation{Department of Physics, Nanjing University, Nanjing 210093, China} 
\affiliation{Center for High Energy Physics, Peking University, Beijing 100871, China} 
\affiliation{Nanjing Proton Source Research and Design Center, Nanjing 210093, China} 
\affiliation{CAS Center for Excellence in Particle Physics, Beijing 100049, China} 

\author{Van Que Tran} 
\affiliation{Department of Physics, Nanjing University, Nanjing 210093, China}

\author{Yilun Xue} 
\affiliation{Department of Physics, Nanjing University, Nanjing 210093, China}

\begin{abstract}

We present a dark matter model to 
explain the excess events in the 
electron recoil data  
recently reported by the 
Xenon1T experiment. 
In our model, dark matter $\chi$ 
annihilates into a pair of on-shell particles $\phi$ 
which subsequently decay into $\psi \psi$ 
final state; $\psi$ interacts with 
electron to generate the observed excess 
events.  
Due to the mass hierarchy, 
the velocity of $\psi$ can be rather large 
and can have an extended distribution, 
which provides a good fit to the electron 
recoil energy spectrum. 
We estimated the flux of $\psi$ 
from dark matter annihilations in the galaxy 
and further determined the interaction 
cross section which is sizable but small enough 
to allow $\psi$ to penetrate the rocks to reach 
the underground labs.   
\end{abstract}

\maketitle

%\section{Introduction}

{\it Introduction:--}Recently, the XENON collaboration reported a new 
result in the low energy electron recoil data 
in the Science Run 1 (SR1) data collected by the Xenon1T experiment 
from February 2017 to February 2018 with an exposure of 
0.65 tonne-years: 
285 events are observed in the electron recoil energy 
between 1 keV and 7 keV, while the background 
expectation is $232 \pm 15$ events \cite{Aprile:2020tmw}. 
Various backgrounds for the excess events are studied 
by the XENON collaboration \cite{Aprile:2020tmw}. 
The uncertainty in the Pb-214 background 
is found not big enough to explain the excess \cite{Aprile:2020tmw}. 
Although the $\beta$ emission of tritium gives a good fit 
to the excess data, the amount of tritium 
due to cosmogenic activation is found to be much smaller than 
needed for the excess \cite{Aprile:2020tmw}. 
But there could be some other sources of tritium 
inside the Xenon1T detector.

Axions produced from the Sun and 
neutrinos with magnetic moment can provide a good fit 
to the excess events with a significance of 
3.5 $\sigma$ and 3.2 $\sigma$  
respectively \cite{Aprile:2020tmw}. 
However, the solar axion explanation is strong 
tension with the stellar cooling constraint 
\cite{Giannotti:2017hny, Viaux:2013lha, Bertolami:2014wua, Battich:2016htm}, 
and neutrinos with magnetic moment are also constrained 
 \cite{Aprile:2020tmw}. 
Axion-like particles \cite{Takahashi:2020bpq}, 
and dark matter (DM) particles with very 
large velocity \cite{ Kannike:2020agf} 
are proposed to explain the excess data.

\begin{figure}[htbp]
\begin{center}
\includegraphics[width=.25\textwidth]{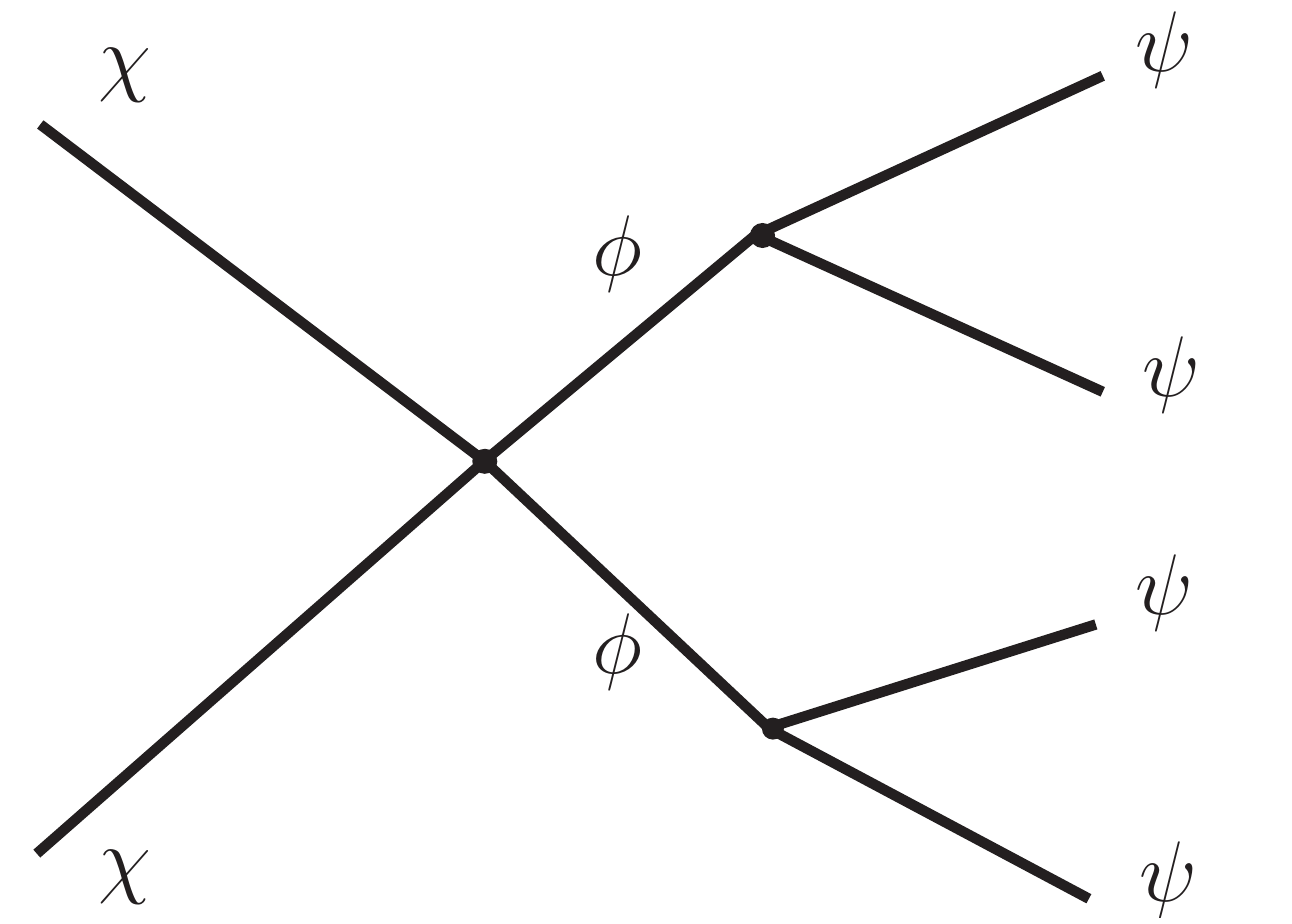}
\caption{Diagrams for the on-shell mediator model.}
\label{diagram}
\end{center}
\end{figure}

In this paper, we present a DM model which 
can explain the excess events in the Xenon1T 
low energy electron recoil data. In this model, 
DM particle $\chi$ annihilates into 
a pair of on-shell particles $\phi$ 
which subsequently decays into 
$\psi \psi$ final state, 
as shown by the diagram on  
Fig.\ (\ref{diagram}). 
We refer to this model as the on-shell mediator model 
(see e.g.\ \cite{Ibarra:2012dw, Mardon:2009rc, Abdullah:2014lla, Cline:2014dwa, Agrawal:2014oha, Cline:2015qha} 
for early studies). 
We assume that $\psi$ has a sizeable interaction 
cross section with electron. 
The produced $\psi$ particle 
can have rather large velocity leading to a $\sim$keV 
electron recoil energy to be recorded by the Xenon1T 
detector.

We show that the $\psi$ particle can have an extended velocity 
distribution due to the mass hierarchy, 
which provides a good fit to the excess spectrum in 
the Xenon1T electron recoil data. 
We further estimate the flux of $\psi$ and 
the interaction cross section with the electron. 
We find that the flux of $\psi$ is consistent with the 
expectation from DM annihilation in the 
galaxy, and the interaction cross section 
needed for the excess is small enough 
such that $\psi$ is not stopped by the 
rock on top of the underground labs.

%\section{On-shell mediator model}

{\it On-shell mediator model:--}We consider a hidden sector 
that contains three particles $\chi$ which is the DM, 
$\phi$, and $\psi$. 
We assume the following mass hierarchy 
$m_\chi > m_\phi > 2 m_\psi$ so that 
DM $\chi$ can annihilate in the following way 
\be
\chi \chi \to \phi \phi \to \psi \psi \psi \psi 
\ee
\footnote{A different mass hierarchy 
can lead to the following process  
$
\chi \chi \to \psi \psi
$
where $\psi$ is mono-energetic.
%This is similar to the scenarios 
%considered in Strumia paper 2006.10735. 
}
Assuming $\psi$ is isotropically produced the 
rest frame of $\phi$, 
the energy spectrum of the $\psi$ particle 
has a box-shape in the energy range 
\be
E_{-} < E_\psi < E_{+}
\ee
where 
$
E_{\pm} = (m_\chi / 2) (1\pm x y)
$
with $x=\sqrt{1-{m_\phi^2 / m_\chi^2}}$ 
and $y=\sqrt{1-{4 m_\psi^2 / m_\phi^2}}$. 
This leads to a velocity distribution of $\psi$ 
as follows 
\be
\int_{v_{-}}^{v_{+}} dv_\psi f(v_\psi) 
=\int_{v_{-}}^{v_{+}} 
\frac{dv_\psi\, v_\psi\,  m_{\psi}}{(E_{+}-E_{-})(1-v_\psi^{2})^{3 / 2}}
\label{eq:fv}
\ee
where $v_\psi(E_\psi)=\sqrt{1-m_\psi^2/E_\psi^2}$ is the velocity of 
$\psi$, and $v_{\pm}=v_\psi(E_{\pm})$.

%\section{Electron recoil spectrum}

{\it Electron recoil events:--}The differential
 rate due to scattering between 
$\psi$ and electron can be computed by 
\cite{Essig:2011nj, Essig:2012yx, Roberts:2016xfw, Roberts:2019chv}
\be
\frac{ d\sigma v_\psi}{dE_R} = \frac{\bar{\sigma}_{e\psi}} {2 m_e} 
\int \frac{dv_\psi f(v_\psi)}{v_\psi} 
\int_{q_{-}}^{q_{+}} a_0^2 qdq |F (q)|^2 K(E_R,q),
\label{eq:diffsigma}
\ee
where $f(v_\psi)$ is the $\psi$ velocity distribution given in Eq.~\ref{eq:fv}, 
$F (q)$ is the dark matter form factor which is  
assumed to be unity in this analysis, $E_R$ is the electron recoil energy, 
$K(E_R,q)$ is the dimensionless atomic excitation factor 
\cite{Roberts:2019chv,Roberts:2016xfw} and
$\bar{\sigma}_{e\psi}$ is the free electron cross section 
at fixed momentum transfer of $q = 1/a_0$ 
with $a_0 = 1/(m_e \alpha)$ being the Bohr radius. 
The integration limits on the momentum transfer is given by 
\cite{Essig:2011nj, Essig:2012yx, Roberts:2019chv}
\be
q_{\pm} = m_\psi v_\psi \pm \sqrt{m_\psi^2 v_\psi^2 - 2 m_\psi E_R}.
\label{qminmax} 
\ee

The differential event rate then can be obtained by 
\be
\frac{dR}{dE_R} = N_T n_\psi \frac{d\sigma v_\psi}{dE_R},
\ee
where $N_T$ is the number of atoms in the target material, 
and $n_\psi$ is the number density of the incident 
$\psi$ particle. We take 
$N_T \simeq 4.2 \times10^{27} \,{\rm ton}^{-1}$ 
for Xenon atoms.

The number of events are then calculated by 
\be
N_{\rm S} = {\rm exposure}\int_{E_1}^{E_2} dE_R \frac{dR}{dE_R} \epsilon (E_R),
\label{nevents}
\ee
where the exposure is 0.65 tonne years, and  $\epsilon (E_R)$ is the total efficiency 
given in the Xenon1T experiment \cite{Aprile:2020tmw}.

%\section{Fitting to excess events}

\begin{figure}[htbp]
\begin{centering}
\includegraphics[width=0.4\textwidth]{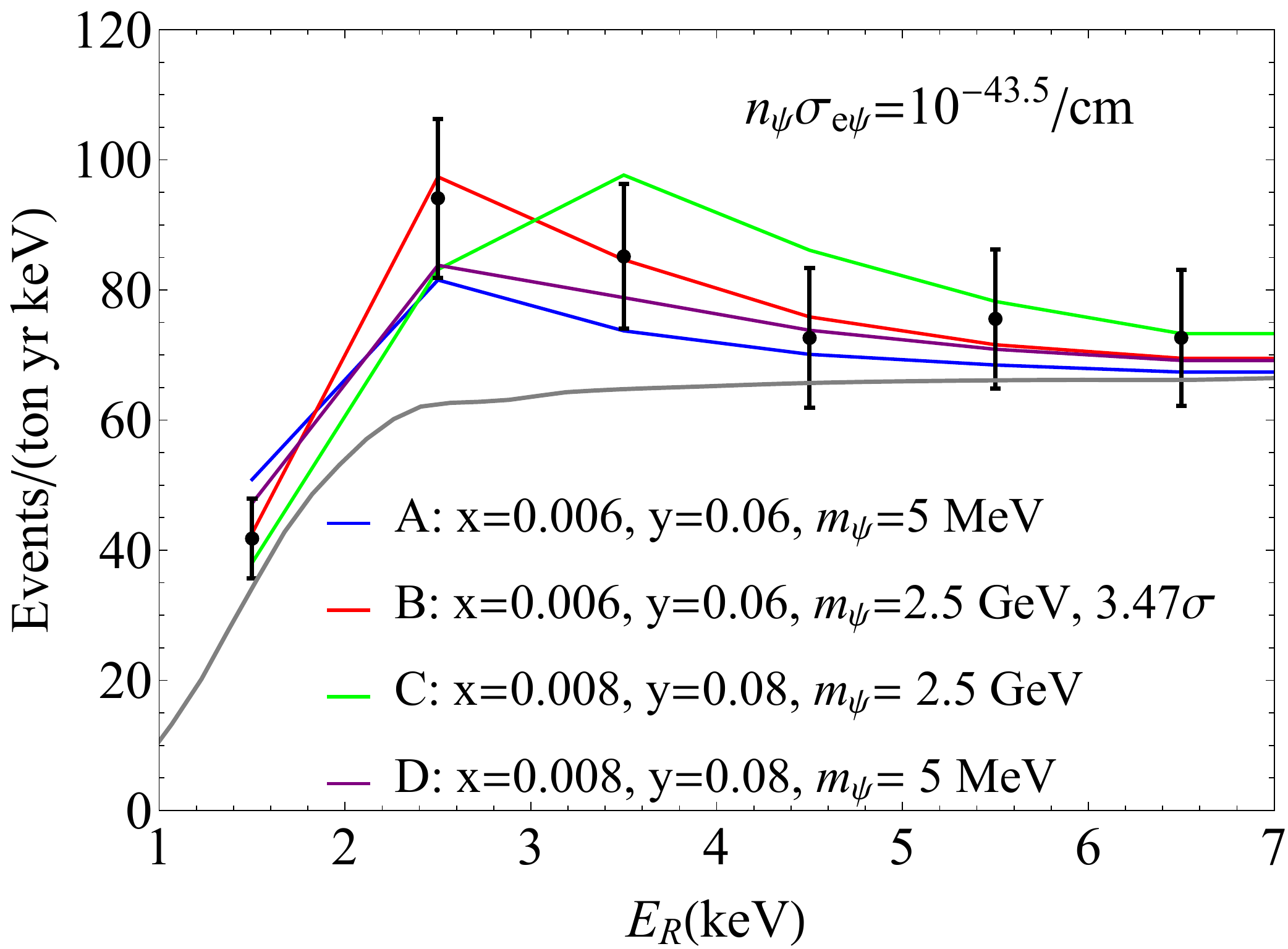} 
\caption{The binned signal events from four 
 benchmark points (A, B, C, D) 
 in the on-shell mediator model.
The black dots are the observed events 
and the gray line is the expected background \cite{Aprile:2020tmw}.
 $n_\psi \sigma_{e\psi} = 10^{-43.5}$/cm 
 is assumed for these benchmark model points. 
}
\label{fig:Nevents}
\end{centering}
\end{figure}

{\it Fitting to excess events:--}We use the on-shell 
mediator models to fit the excess events 
in the electron recoil energy range 1-7 keV on 
Fig.~\ref{fig:Nevents}. 
We take the background from the Xenon1T 
paper \cite{Aprile:2020tmw}. 
We considered four benchmark model points 
on Fig.~\ref{fig:Nevents}. 
The product of the number density of $\psi$ and 
its interaction cross section with electron
is fixed to be $n_\psi \sigma_{e\psi} = 10^{-43.5}$/cm 
for all benchmark model points on Fig.~\ref{fig:Nevents}.

We carry out a simple chi-square analysis by 
analyzing only the six low energy electron 
recoil data points shown on Fig.~\ref{fig:Nevents},  
where $\chi^2$ is calculated by 
\be
\chi^2 = \sum_i \frac{(N^i_{\rm exp} - N^i_{\rm th})^2}{(\delta N^i_{\rm exp})^2}
\ee
where $i$ denotes for bins, 
$N^i_{\rm exp}$ and $\delta N^i_{\rm exp}$ are the number of observed events 
and its uncertainty taken from Xenon1T \cite{Aprile:2020tmw}. 
$N^i_{\rm th}$ 
is the number of signal events calculated in Eq. \ref{nevents} 
plus the expected background taken 
from Xenon1T \cite{Aprile:2020tmw}.  
It is found that the benchmark model point B 
has the best-fit significance of $3.5 \sigma$ 
among the four points; the 
benchmark model points 
A, C, D yields a significances of 
$2.7\sigma$, $3.0\sigma$ and $3.2\sigma$
respectively. 

Also, comparing the BP B and C, 
it is found that the peak bin tends to shift to the right 
when $x$ and $y$ have bigger values.  
Lower masses BPs result in a smaller number of events 
in the excess region.

%\section{Scan}

%{\it Scan in the parameter space:--}
We further 
carry out a scan in the projected parameter space spanned 
by $m_\psi$ and $(m_\chi - m_\phi)$, both in 
the range of 0.001 GeV to 10 GeV. 
The relation $m_{\phi} = 2 m_{\psi}+0.1$ MeV 
is used in in the scan. 
The model points equally distributed on both dimensions in the 
log scale.

\begin{figure}[htbp]
\begin{center}
\includegraphics[width=.4\textwidth]{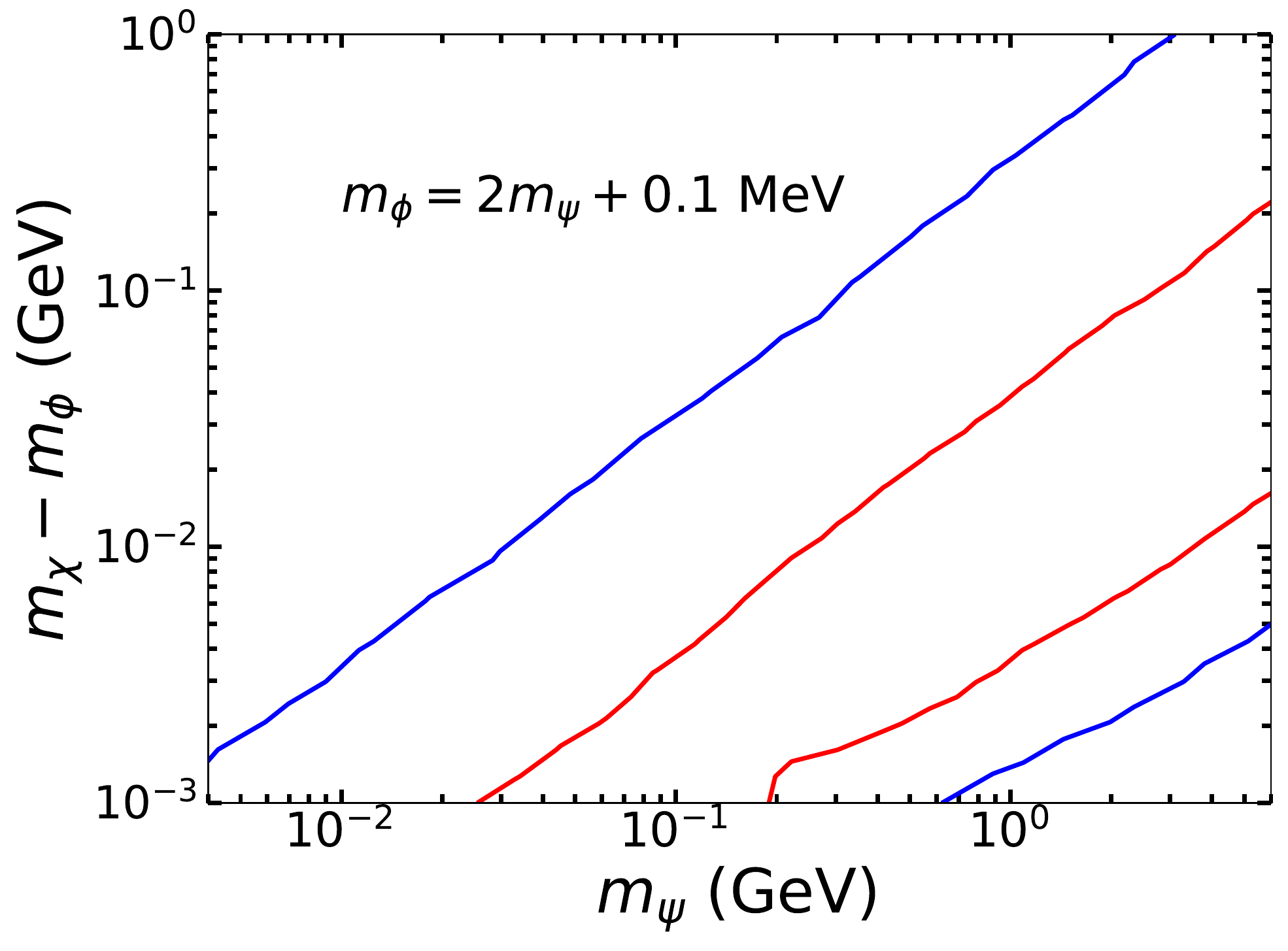}
\caption{The Xenon1T excess fitting region on the plane of 
the $m_\psi$ and the mass gap $m_\chi- m_\phi$. 
Here, we fixed $m_{\phi} = 2 m_{\psi}+0.1$ MeV.
The red (blue) contour represents $68\%$ ($95\%$) C.L. 
}
\label{fig:scan}
\end{center}
\end{figure}

Fig. \ref{fig:scan} shows the fit to 
the Xenon1T excess as function of the $m_\psi$ and $(m_\chi - m_\phi)$.
It is found that $\chi^2 = 1.2$ for the best-fit model point 
at $(m_{\psi}, m_{\chi} - m_{\phi})$ $= (0.99, 0.009)$ GeV.
The red and blue contours represent the $68\%$ and $95\%$ C.L. 
which, in a two dimensional parameter space, corresponds 
to $\Delta \chi^2=2.3$ and $\Delta \chi^2=$5.99 respectively.

For smaller mass gap between $\chi$ and $\phi$ particles, 
the final state particle $\psi$ has higher velocity; 
thus in order to produce enough excess events 
in the small electron recoil energy range (1-7 keV), 
smaller mass values of $\psi$ are needed. 
This result in a tendency shape in Fig. \ref{fig:scan}.
We found that in order to generate the Xenon1T excess,
the $\psi$ velocity has to be in the range of $v_\psi/c \sim (0.01, 0.1)$.

%\section{Particle flux of $\psi$}
{\it Particle flux of $\psi$:--}To compute the flux of $\psi$ 
from DM annihilations, 
we assume an NFW profile for the Milky Way DM halo 
\be
\rho_\chi(r)=\rho_{s} \frac{\left(r / r_{s}\right)^{-\gamma}}{\left(1+r / r_{s}\right)^{3-\gamma}}
\ee
where we take $\gamma =1$, 
$\rho_{s} =0.31$ GeV/cm$^3$, 
and $r_{s} =21$ kpc. 
The flux of $\psi$ is given by 
\be
\Phi_{\psi}=4 \frac{\langle\sigma v\rangle}{8 \pi m_{\chi}^{2}} J
\ee
where the total J-factor is $J=\int d \Omega \int d s \rho_{\chi}^{2} 
\simeq  10^{23}$ $\mathrm{GeV^2/cm^5}$, 
and $\langle\sigma v\rangle$ is the DM annihilation cross section. 
The total $\psi$ flux is $\Phi_{\psi} \simeq 10^{-3}$ cm$^{-2}$ s$^{-1}$  
for $m_{\chi} \simeq$ GeV if the canonical thermal cross section 
$\langle\sigma v\rangle = 3\times 10^{-26} \mathrm{cm^3/s}$ 
is assumed. 
We note the particle flux of $\psi$ can be further enhanced if there exists 
some DM subhalos in the vicinity of the solar system.

Using the benchmark model points B and C on Fig.~\ref{fig:Nevents}, 
we determine the interaction cross section to be 
$\sigma_{\psi e} \simeq 10^{-32}$ cm$^2$. 
This is much larger than the dark matter direct detection 
upper limit  $\sigma_{\rm DM-e} \lesssim10^{-38}$ cm$^2$ \cite{Essig:2017kqs}. 
However, the particle flux of $\psi$ is about 8 order of magnitude smaller than 
the local DM flux $\Phi_{\chi} \simeq$ 
10$^{5}$ cm$^{-2}$ s$^{-1}$ 
for $m_\chi \simeq$ GeV. 
Thus $\psi$ with 
$\sigma_{\psi e} \simeq 10^{-32}$ cm$^2$ 
is allowed by DM direct detection limits.

However, if the interaction cross section between 
$\psi$ and electron is so large that it is absorbed 
by the rock on top of the underground labs.  
In order to reach the underground labs, 
the interaction cross section has to satisfy 
$\sigma_{\psi e} \lesssim 10^{-23}$ cm$^{-2}$  
\cite{Emken:2019tni}. 
Thus $\psi$ with 
$\sigma_{\psi e} \simeq 10^{-32}$ cm$^2$ 
can enter the Xenon1T to generate the 
excess events.

%\section{Discussion}

{\it Discussion:--}A possible realization of the on-shell mediator 
model discussed is a hidden sector model 
in which $\chi$ and $\psi$ are fermions charged 
under the hidden $U(1)$ 
gauge boson $\phi = \phi_\mu$. 
Both $\chi$ and $\psi$ are stable due to 
the  hidden $U(1)$. 
The interaction Lagrangian is given by 
\be
g_h \phi_\mu (\bar \chi \gamma^\mu \chi  
+  \bar \psi \gamma^\mu \psi)
%+ \epsilon e \bar f \gamma^\mu f  ) 
\ee
$\psi$ can either interact with electron via 
some electrophilic interaction, or 
interact with the standard model fermions 
via another gauge boson $A'_\mu$ which 
is kinetically mixed with the standard model 
hypercharge boson. 
We note that such models 
can be searched for in $e^+e^-$ 
colliders.

The mass hierarchy 
$m_\chi > m_\phi > 2 m_\psi$ leads to 
the box-shape energy spectrum of $\psi$. 
If $m_\phi > m_\chi$, the annihilation 
process $\chi\chi \to \phi \to \psi\psi$ 
dominates; the energy spectrum of 
$\psi$ is a delta function, smeared by 
the small kinetic energy of DM $\chi$. 
The velocity distribution in the 
$\chi\chi \to \phi \to \psi\psi$ case has 
been investigated in Ref.\ \cite{Kannike:2020agf}.

%\section{Conclusion}

{\it Conclusion:--}We have analyzed the on-shell mediator DM models 
to fit the excess events in the 
low energy electron recoil data 
observed by Xenon1T. 
We find that the 
on-shell mediator DM models  
can provide a good fit to the Xenon1T data. 
The benchmark models that can explain 
the Xenon1T excess are consistent with 
the expected flux arising from DM 
annihilations in the Milky Way DM halo. 
Although the interaction cross section 
needed for the excess is sizable, 
it is small enough such that $\psi$ 
can penetrate the rock to reach the 
underground labs.

%\section{Acknowledgement}

%%
{\it Acknowledgement:--}We thank Ma, Yue 
for discussions 
and correspondence. 
The work is supported in part  
by the National Natural Science 
Foundation of China under Grant Nos.\ 
U1738134 and 11775109.

\end{document}